\input{aipcheck}

\documentclass[
    ,final            
  ]
  {aipproc}

\layoutstyle{6x9}

\def \inte {{\it INTEGRAL\,}}
\def \xmm {{\it XMM--Newton}}

\def \sw {{\it Swift}}

\def \hcm {\hbox {\ifmmode $ atom cm$^{-2}\else atom cm$^{-2}$\fi}}

\def \ATel {Astron.\ Tel.}
\def \apj {ApJ}
\def \apjl {ApJL}

\def \aap {A\&A}

\def \pasj {PASJ}

\def \mnras {MNRAS}

\begin{document}

\title[The Swift view of Supergiant Fast X--ray Transients]{The Swift view of Supergiant Fast X--ray Transients}


\author{L. Sidoli}{
  address={INAF, Istituto di Astrofisica Spaziale e Fisica Cosmica, Via E. Bassini 15, I-20133 Milano, Italy}
}

\author{P. Romano}{
  address={INAF, Istituto di Astrofisica Spaziale e Fisica Cosmica, Via La Malfa 153, I-90146 Palermo, Italy}
}

\author{L. Ducci}{
  address={Universit\`a degli Studi dell'Insubria, Via Valleggio 11, I-22100 Como, Italy}
  ,altaddress={INAF, Istituto di Astrofisica Spaziale e Fisica Cosmica, Via E. Bassini 15, I-20133 Milano, Italy} 
}

\author{A. Paizis}{
  address={INAF, Istituto di Astrofisica Spaziale e Fisica Cosmica, Via E. Bassini 15, I-20133 Milano, Italy}
}

\author{S. Vercellone}{
  address={INAF, Istituto di Astrofisica Spaziale e Fisica Cosmica, Via La Malfa 153, I-90146 Palermo, Italy}
}

\author{G. Cusumano}{
  address={INAF, Istituto di Astrofisica Spaziale e Fisica Cosmica, Via La Malfa 153, I-90146 Palermo, Italy}
}

\author{V. La Parola}{
  address={INAF, Istituto di Astrofisica Spaziale e Fisica Cosmica, Via La Malfa 153, I-90146 Palermo, Italy}
}

\author{V. Mangano}{
  address={INAF, Istituto di Astrofisica Spaziale e Fisica Cosmica, Via La Malfa 153, I-90146 Palermo, Italy}
}

\author{D.N.\ Burrows}{
  address={Department of Astronomy and Astrophysics, Pennsylvania State  University}
}

\author{J.A.\ Kennea}{
  address={Department of Astronomy and Astrophysics, Pennsylvania State  University}
}

\author{H.A.\ Krimm}{
  address={CRESST/Goddard Space Flight Center, Greenbelt, MD, USA \\
Universities Space Research Association, Columbia, MD, USA}
 ,altaddress={NASA/Goddard Space Flight Center, Greenbelt, MD 20771, USA}
}

\author{N.\ Gehrels}{
  address={NASA/Goddard Space Flight Center, Greenbelt, MD 20771, USA}
}

\begin{abstract}
We report here on the recent results of a monitoring campaign 
we have been carrying out with $Swift$/XRT on a sample of
four Supergiant Fast X--ray Transients. 
The main goal of this large programme 
(with a net $Swift$/XRT exposure of $\sim$540~ks, updated to 2009, August, 31) 
is to address several main open
issues related to this new class of High Mass X--ray Binaries (HMXBs)
hosting OB supergiant stars as companions.
Here we summarize the most important results obtained 
between October 2007 and  August 2009.
\end{abstract}

\maketitle


\section{Supergiant Fast X--ray Transients}

The discovery with the \inte\ satellite 
of a new class of X--ray Binaries (the Supergiant Fast X--ray Transients, SFXTs) 
associated with blue supergiant companions and 
characterized by fast transient X--ray emission \cite{Sguera2005}, has changed the classical view of 
the HMXBs research field.
This new subclass of massive binaries includes 8 members, with $\sim$20 candidates 
(see \cite{Bird2009b} or  {\em http://isdc.unige.ch/$\sim$rodrigue/html/igrsources.html} 
for a list of new \inte\ transient sources).
These sources show apparently short X--ray outbursts (as observed with \inte), 
characterized by a few hours duration flaring activity, reaching 10$^{36}$--10$^{37}$~erg~s$^{-1}$ at peak,
but with no persistent emission, as observed with IBIS/\inte.

Follow-up observations allowed to identify the  companions as 
OB supergiant stars (e.g. \cite{Pellizza2006, Masetti2006, Negueruela2006}).

A few sources  were observed with more sensitive instruments ($Chandra$ and \xmm) outside
bright flares, being found in a quiescent state, with a very soft, likely thermal, spectrum
and a low X--ray luminosity of  $\sim$10$^{32}$~erg~s$^{-1}$ (e.g. IGR~J17544--2619, \cite{zand2005}).

Spectral properties seem to be very similar to those of accreting pulsars, with a flat power law spectrum below
10~keV, and high energy cutoff around 10--30 keV (e.g. \cite{Smith1998:17391-3021, SidoliPM2006}).

Thus, it is often assumed that all SFXTs host a neutron star as compact object, although only in four members 
a pulse period has been measured:  
AX~J1841.0$-$0536 ($P_{\rm spin}\sim4.7$\,s, \cite{Bamba2001}), 
IGR~J11215--5952 ($P_{\rm spin}\sim187$\,s, \cite{Swank2007}), 
IGR~J16465--4507  ($P_{\rm spin}\sim228$\,s, \cite{Lutovinov2005}) and 
IGR~J18483--0311 ($P_{\rm spin}\sim21$\,s, \cite{Sguera2007}).
A black hole cannot be excluded in the other members of the class.

The orbital periods have been measured in five sources and range
from 3.3~days (IGR~J16479--4514, \cite{Jain2009:16479}) 
to 165~days (in IGR~J11215--5952, \cite{SidoliPM2006, Sidoli2007, Romano2007, Romano2009:11215_2008}).
Two sources display periodically recurrent outbursts,  IGR~J11215--5952 \cite{SidoliPM2006} 
and IGR~J18483--0311 (\cite{Levine2006, Sguera2007}), 
suggesting that the outburst is triggered near the periastron
passage in a highly eccentric orbit.

\section{$Swift$ contribution}

We performed the first 
long-term monitoring campaign with \sw/XRT of a sample of four SFXTs:
XTE~J1739--302, IGR~J17544--2619 (the two prototypes of the class),
IGR~J16479--4514 and the X--ray pulsar AX~J1841.0$-$0536.

The campaign strategy consists of 2 or 3 XRT pointings per source per week (about $\sim$1~ks each)
in order to monitor the source status, which was largely unknown (outside the bright outbursts) before 
these \sw\ observations. 
Given the structure of the observing plan, this monitoring can be considered as a 
casual sampling of the source light curves at a resolution of about $\sim 4$\,days (see Table~1 for a summary). 

The main aims were to monitor the long-term properties, 
to catch ``almost'' every outburst (even the fainter ones, 
not triggering the \sw/BAT), 
to monitor the onset of a new outburst and to be able, in this case, 
to follow the whole outburst duration with
more frequent subsequent observations (in this respect the \sw\ flexibility is a crucial property),
and to get truly simultaneous wide band spectra during bright flares.
Our monitoring campaign (still on-going) unveiled several new properties of SFXTs 
(see details in \cite{Sidoli2008:sfxts_paperI,
Romano2008:sfxts_paperII,
Sidoli2009:sfxts_paperIII,
Sidoli2009:sfxts_paperIV, Romano2009:sfxts_paperV}).

In particular, \sw\ observations have demonstrated that SFXTs spend most 
of their life still accreting matter even outside bright flaring 
activity, emitting at an intermediate level of 10$^{33}$--10$^{34}$~erg~s$^{-1}$, 
with large variability and an absorbed  hard X--ray
spectrum (power law photon index of 1--2, or hot black body temperatures of 1--2~keV, observed with XRT
below 10 keV).

Besides the bright outbursts (detected also with BAT) and the intermediate level of X--ray
emission, several 3$\sigma$ upper limits were also obtained, either because the source was faint or because
of a short exposure due to the interruption by gamma-ray burst events. 
Thus, to create a uniform subsample 
for the ``non-detections'' state, we excluded all observations that had a net exposure below 900\,s.
An exposure of 900\,s corresponds to 2--10\,keV flux limits 
of $\sim$1--3$\times 10^{-12}$ erg cm$^{-2}$ s$^{-1}$ (3$\sigma$), 
depending on the source (assuming the best fit absorbed power law model 
for the intermediate state of each source). 

We then defined as {\it duty cycle of inactivity} (IDC), 
the time fraction each source spends undetected down to a flux limit of 
1--3$\times10^{-12}$ erg cm$^{-2}$ s$^{-1}$ (which means an upper limit to the time spent in quiescence). 
The IDCs are the following: 17~\% (IGR~J16479--4514), 28~\% (AX~J1841.0--0536), 39~\% (XTE~J1739--302) 
and 55~\% (IGR~J17544--2619).

For IGR~J16479$-$4514 a contribution to the IDC is due to the X--ray eclipses. 
One of the main findings of our monitoring is that the quiescence in these transients 
is a rarer state \cite{Romano2009:sfxts_paperV} than what previously thought 
based only on \inte\ observations.  
Accumulating all data for which no detections were obtained as single exposures \cite{Romano2009:sfxts_paperV}, we could get 
the lowest luminosity level, which is reached in XTE~J1739--302 (6$\times$10$^{32}$~erg~s$^{-1}$, 2--10 keV) and 
in IGR~J17544$-$2619 (3$\times$10$^{32}$~erg~s$^{-1}$).
 
Broad band truly simultaneous spectra (XRT together with BAT) were obtained from 8 bright flares
caught from 3 of the 4 monitored sources. 
The best fits could be obtained with Comptonized models ({\sc compTT} or {\sc bmc} in {\sc xspec}) 
or with an absorbed flat power law model with
high energy cutoff around 10--30 keV 
(see \cite{Romano2008:sfxts_paperII, Sidoli2009:sfxts_paperIII, Sidoli2009:sfxts_paperIV}). 
This spectral shape, and in particular the spectral cutoff, is compatible with a neutron star magnetic field
of $\sim$10$^{12}$~G \cite{Coburn2002}, although no cyclotron lines have been detected
yet in the SFXTs we have monitored (but see also \cite{Sguera2010} for a hint of a cyclotron emission line at low energy
in the SFXT IGR~J18483--0311).

Variable absorbing column densities were also observed between different outbursts
in the same source (XTE~J1739--302) and within the same outburst, likely due 
to local dense clouds of matter composing the supergiant wind.

The SFXTs bright and short flares (a few hour long) are actually 
part of a longer outburst phase lasting days \cite{Sidoli2009:sfxts_paperIII}, 
as already observed in the SFXT IGR~J11215--5952 \cite{Romano2007}.

Optical/UV observations obtained with UVOT
simultaneously to \sw/XRT monitoring  have revealed a possible hint (which needs confirmation) 
of an UV flaring activity simultaneously to the
X--ray bright flares in XTE~J1739$-$302 \cite{Romano2009:sfxts_paperV}.

\begin{table}
\begin{tabular}{lrrrr}
\hline
  & \tablehead{1}{r}{b}{Number of \\ Observations }
  & \tablehead{1}{r}{b}{XRT net \\ exposure (ks) }
  & \tablehead{1}{r}{b}{Number of \\ bright flares}
  & \tablehead{1}{r}{b}{Inactivity \\ Duty Cycle (IDC)}   \\
\hline
IGR~J16479--4514 & 133 & 152 &  2   & 17 \% \\
XTE~J1739--302   & 162 & 165 &  3   & 39 \% \\
IGR~J17544--2619 & 138 & 127 &  4   & 55 \% \\
AX~J1841.0--0536 &  88 & 96.5 & 0   & 28 \% \\
\hline
\end{tabular}
\caption{Status of the \sw/XRT monitoring programme of SFXTs (updated to 2009, August 31)}
\label{tab:a}
\end{table}


\begin{theacknowledgments}
This work was supported in Italy by ASI contracts  I/023/05/0,
I/088/06/0 and I/008/07/0. 
\end{theacknowledgments}

\bibliographystyle{aipproc}   

\IfFileExists{\jobname.bbl}{}
 {\typeout{}
  \typeout{******************************************}
  \typeout{** Please run "bibtex \jobname" to optain}
  \typeout{** the bibliography and then re-run LaTeX}
  \typeout{** twice to fix the references!}
  \typeout{******************************************}
  \typeout{}
 }

\end{document}